\documentclass[a4paper]{ifacconf}

\usepackage{graphicx,amsmath,url}      
\usepackage[round]{natbib}             
\usepackage{ae}
\usepackage{amssymb}
\graphicspath{ {imagens/} }
\usepackage{float}
\usepackage{epstopdf}
\usepackage{color}
\usepackage[dvipsnames]{xcolor}
\usepackage{tensor}
\usepackage{multirow}
\usepackage[normalem]{ulem}
\usepackage{physics}

\def\portugues{1} 

\ifx\portugues\undefined
\def\portugues{0}
\fi

\if\portugues0
   \usepackage[english]{babel}
  \else
   \usepackage[spanish,brazil,english]{babel}
\fi

\usepackage[T1]{fontenc}

\usepackage[utf8]{inputenc}

\usepackage{ae}

\if\portugues1
 { }
 { }
 { }
 
 \newtheorem{exemplo}{{\bf Examplo}}{}

{}
\fi

\begin{document}

\if\portugues1

%
\selectlanguage{brazil}
	
\begin{frontmatter}

\title{Identificação de Sistemas Não Lineares Utilizando o Algoritmo Híbrido e Binário de Otimização por Enxame de Partículas e Busca Gravitacional} 

\thanks[footnoteinfo]{}

\author[First]{Wilson Rocha Lacerda Junior} 
\author[First]{Samir Angelo Milani Martins} 
\author[First]{Erivelton Geraldo Nepomuceno}

\address[First]{GCOM - Grupo de Controle e Modelagem, Departamento de Engenharia Elétrica, Universidade Federal de São João del-Rei, MG, (e-mails: wilsonrljr@outlook.com, martins@ufsj.edu.br, nepomuceno@ufsj.edu.br).}

\selectlanguage{english}
\renewcommand{\abstractname}{{\bf Abstract:~}}
\begin{abstract}                
This work presents a new meta-heuristic approach to model structure selection of polynomial NARX models. In this respect, the technique penalizes the models based on the individual contribution of each regressor in representing the system. The new algorithm is tested on two experimental case studies: the identification of an electromechanical system and a eletric heater. The results are compared with Error Reduction Ratio and another meta-heuristic approach. The proposed method shows its advantages over compared methods in terms of the trade-off between prediction accuracy and model interpretability. The results are quantified and compared using the Mean Squared Error (MSE) indices.

\vskip 1mm
\selectlanguage{brazil}
{\noindent \bf Resumo}:  Esse trabalho apresenta uma nova abordagem baseada em meta-heurística para seleção de estruturas de modelos NARX polinomiais. Nesse sentido, a técnica penaliza os modelos considerando a contribuição individual de cada regressor ao representar o sistema. O novo algoritmo é testado em dois estudos de caso experimentais: a identificação de um sistema eletromecânico e de um aquecedor elétrico. Os resultados são comparados com a Taxa de Redução de Erro e com uma outra técnica baseada em meta-heurística. O método proposto mostra suas vantagens em relação aos métodos comparados em termos do compromisso entre a precisão da predição e a interpretabilidade do modelo. Os resultados são quantificados e comparados por meio do índice Erro Quadrático Médio (MSE).
\end{abstract}

\selectlanguage{english}


\selectlanguage{brazil}

\end{frontmatter}
\else
%

\begin{frontmatter}

\title{Style for SBA Conferences \& Symposia: Use Title Case for
  Paper Title\thanksref{footnoteinfo}} 

\thanks[footnoteinfo]{Sponsor and financial support acknowledgment
goes here. Paper titles should be written in uppercase and lowercase
letters, not all uppercase.}

\author[First]{First A. Author} 
\author[Second]{Second B. Author, Jr.} 
\author[Third]{Third C. Author}

\address[First]{Faculdade de Engenharia Elétrica, Universidade do Triângulo, MG, (e-mail: autor1@faceg@univt.br).}
\address[Second]{Faculdade de Engenharia de Controle \& Automação, Universidade do Futuro, RJ (e-mail: autor2@feca.unifutu.rj)}
\address[Third]{Electrical Engineering Department, 
   Seoul National University, Seoul, Korea, (e-mail: author3@snu.ac.kr)}
   
\renewcommand{\abstractname}{{\bf Abstract:~}}   
   
\begin{abstract}                
These instructions give you guidelines for preparing papers for IFAC
technical meetings. Please use this document as a template to prepare
your manuscript. For submission guidelines, follow instructions on
paper submission system as well as the event website.
\end{abstract}


\end{frontmatter}
\fi


\section{Introdução}
A seleção da estrutura de modelos é um tópico que, em um sentido mais amplo, relaciona-se a muitos temas práticos como, por exemplo, ajuste de dados, previsão de séries temporais, seleção de características em classificação e redução de complexidade em redes neurais \citep{wei2008model, billings2013nonlinear}. Entre as diversas classes de modelos, tem-se que a representação polinomial não linear autorregressiva com entrada exógena (NARX, do inglês Nonlinear AutoRegressive model with eXogenous input), incluindo suas variações, são as mais populares para a identificação de sistemas dinâmicos não lineares no domínio de tempo discreto \citep{billings2013nonlinear}.

A etapa de seleção de estruturas representa o maior desafio na obtenção de modelos NARX polinomiais capazes de representar um sistema adequadamente \citep{billings2013nonlinear, falsone2015randomized, akinola2019non}. A identificação de sistemas frequentemente pode ser interpretada como um problema de otimização em que se busca encontrar o modelo ótimo em um espaço de busca formado por uma grande quantidade de modelos candidatos. Dessa forma, problemas relacionados à sobreparametrização e mau condicionamento numérico são tipicamente provenientes da ineficiência de determinados algoritmos em selecionar, de forma apropriada, quais termos devem compor o modelo final \citep{aguirre1995dynamical, piroddi2003identification}.


Um método clássico tradicionalmente utilizado para seleção de estruturas é sugerido por \cite{korenberg1988orthogonal}, baseado nos Mínimos Quadrados Ortogonais (MQO) e na Taxa de Redução de Erro (do inglês Error Reduction Ratio - ERR). No entanto, devido a estratégia incremental e outras limitações desse método como, por exemplo, sistemas excitados por entradas pouco variantes, muitas vezes soluções sub-ótimas são geradas \citep{falsone2015randomized}. Sendo assim, diversas técnicas tem sido desenvolvidas para tentar solucionar o problema de identificação de sistemas dinâmicos não lineares \citep{guo2015iterative, severino2017meta, tang2018bayesian}.

Apesar do esforço no desenvolvimento de novas técnicas, esses métodos ainda possuem limitações evidentes. A utilização de meta-heurísticas propostas em \citep{severino2017meta}, por exemplo, utiliza o Critério de Informação de Akaike (AIC) para compor a função custo do problema de otimização. De acordo com \cite{chen2003sparse}, esse tipo de critério pode resultar em modelos sobreparametrizados em um contexto não linear. Ainda nesse sentido, o método não verifica a relação entre os regressores dos modelos testados. Além disso, a técnica necessita que se estabeleça três parâmetros de forma empírica: i) valor máximo para a função custo; ii) determinação de um número máximo de termos para compor o modelo final \citep{severino2017meta} e; iii) número de iterações do algoritmo. Em \citep{severino2017meta}, os testes realizados foram comparados com modelos já estabelecidos na literatura, de forma que esse tipo de escolha poderia ser bem definida. Entretanto, em uma situação prática, nenhuma dessas informações é conhecida \textit{a priori}, o que torna o procedimento limitado em um contexto em que não há nenhuma informação sobre o sistema a ser modelado. 

À vista do exposto, esse artigo propõe uma técnica de identificação de sistemas não lineares utilizando meta-heurísticas que não depende de nenhuma restrição inicial em relação ao modelo desejado, ou seja, não depende da pré-determinação do número máximo de termos ou do valor máximo para a função custo. Para isso, é proposta uma nova alternativa para compor a função custo do problema, na qual se verifica a importância de cada regressor na composição do modelo a fim de se obter a melhor solução global. Além disso, uma nova forma de codificação é utilizada ao utilizar um algoritmo binário composto pela combinação de duas meta-heurísticas, permitindo um melhor desempenho ao buscar a solução ótima do problema.

O artigo está estruturado da seguinte forma: a presente seção apresentou uma breve introdução histórica sobre o tema a ser desenvolvido. Na seção 2 são levantados os conceitos preliminares, suficientes para o entendimento do trabalho. Na seção 3 são apresentados os métodos utilizados para se chegar aos resultados discutidos na seção 4. A seção 5 trata da conclusão do trabalho e propostas de possíveis trabalhos futuros.

\section{Conceitos Preliminares}

\subsection{Identificação de Sistemas}

As principais etapas do processo de identificação de sistemas consistem em \citep{aguirre}: 
\begin{enumerate}
    \item Testes dinâmicos e coleta de dados; dados experimentais do sistema a ser identificado são coletados, sendo imprescindível o projeto adequado do sinal de excitação a ser utilizado, bem como a frequência de amostragem dos dados coletados. 
    \item Escolha da representação matemática a ser usada; 
    \item Determinação da estrutura do modelo; O algoritmo mais utilizado para seleção de estruturas de modelos é a Taxa de Redução de Erro em conjunto com o Critério de Informação de Akaike \citep{wei2008model}. Devido sua facilidade de implementação e eficiência computacional, esse método se tornou padrão para aproximações de funções lineares e treinamento de redes neurais \citep{haykin2009neural}. 
    
    \item Estimação de parâmetros; os parâmetros estimados quantificam a contribuição de cada regressor em um dado modelo. Uma vez que os modelos NARX polinomiais são lineares nos parâmetros, técnicas de estimação de parâmetros baseadas no método dos mínimos quadrados são frequentemente utilizadas.
    
    \item Validação; essa etapa consiste em verificar se um modelo é adequado ou não para determinada aplicação. Para quantificar esta validação, índices como o Erro Quadrático Médio (MSE) \citep{aguirre} podem ser utilizados. O MSE é definido como:
    \begin{equation}
             MSE= \frac{1}{N}\sum_{i=1}^{N} (y-\hat{y})^2 ,
    \end{equation}
\noindent sendo $\hat{y}(k)$ a saída do modelo e $y(k)$ a saída medida. 
\end{enumerate}

\subsection{Modelos NARX polinomiais}
Um modelo NARX polinomial pode ser definido como \citep{chen1989representations}:
\begin{align}
             y(k)=& F^\ell[y_{k-1},...,y_{k-n_y}\nonumber \\
              &u_{k-d},...,u_{k-d-n_u}] + e_k, 
             \label{eqnarx}
\end{align}
\noindent em que $n_y$, $n_u$, $d$, $u_k$ e $y_k$ são os atrasos em $y$, em $u$, o tempo morto, o sinal de entrada e o sinal de saída no instante $k$, respectivamente. $e_k$ representa ruído e possíveis incertezas que não podem ser bem representados por $F^\ell$, que é uma função polinomial de $y_k$ e $u_k$ com grau de não linearidade $\ell \in \mathbb{N}$. Em uma forma mais compacta, o modelo descrito pela Equação~\eqref{eqnarx} pode ser representado na forma de matriz como:
\begin{equation}
\label{eq5:narx_matrix_form}
    y = \Psi\hat{\Theta} + \Xi,
\end{equation}
\noindent em que $\Psi$ é um vetor formado pela combinação de regressores, $\hat{\Theta}$ é o vetor de parâmetros estimados e $\Xi$ é o vetor de resíduos, representados como:
\begin{align}
\label{eq5:narx_matrix_form_elements}
     y = \begin{bmatrix}
    y_1 \\
    y_2 \\
    \vdots \\
    y_n 
    \end{bmatrix},
    \Psi = \begin{bmatrix}
    \psi_{\indices{_1}} \\
    \psi_{\indices{_2}} \\
    \vdots \\
    \psi_{\indices{_{n_{\Theta}}}}
    \end{bmatrix}^\top=
    \begin{bmatrix}
    \psi_{\indices{_1}1} & \psi_{\indices{_2}1} & \dots & \psi_{\indices{_{n_{\Theta}}}1} \\
    \psi_{\indices{_1}2} & \psi_{\indices{_2}2} & \dots & \psi_{\indices{_{n_{\Theta}}}2} \\
    \vdots & \vdots &       & \vdots \\
    \psi_{\indices{_1}n} & \psi_{\indices{_2}n} & \dots & \psi_{\indices{_{n_{\Theta}}}n} \\
    \end{bmatrix}, \nonumber \\
    \hat{\Theta} = \begin{bmatrix}
    \hat{\Theta}_1 \\
    \hat{\Theta}_2 \\
    \vdots \\
    \hat{\Theta}_{n_\Theta} 
    \end{bmatrix},
    \Xi = \begin{bmatrix}
    \xi_1 \\
    \xi_2 \\
    \vdots \\
    \xi_n 
    \end{bmatrix}.
\end{align}

O número máximo de termos candidatos durante o processo de seleção de estrutura do modelo pode ser definido como \citep{korenberg1988orthogonal}:
\begin{equation}
    \label{n1}
    n_{\Theta} = M+1,
\end{equation}

\noindent em que
\begin{align}
\label{ni}
    M = & \sum_{i=1}^{\ell}n_i  \nonumber \\
    ni = & \frac{n_{i-1}(n_y+n_u+i-1)}{i}, n_{0} = 1. 
\end{align}

Um modelo (Equação~\eqref{eqnarx}) contendo todos os termos gerados considerando $n_y$, $n_u$ e $\ell$ pode facilmente se tornar sobreparametrizado, como mostram as Equações~\eqref{n1} e \eqref{ni}. Um modelo dinâmico de terceira ordem ~($n_y = n_u = 3$) expandido como um polinômio cúbico ($\ell = 3$) teria, por exemplo, $84$ termos. Aumentando-se a ordem, o grau de não linearidade e o número de entradas e saídas do sistema, o número de termos candidatos cresce significantemente. Sabe-se, ainda, que há termos que não são relevantes para o modelo, o que evidencia a importância da utilização de uma boa técnica de seleção de estruturas.

\subsection{BPSOGSA - Algoritmo de otimização híbrido e binário utilizando Otimização por Enxame de Partículas (PSO) e Busca Gravitacional (GSA)}

O algoritmo híbrido BPSOGSA foi introduzido por \citet{mirjalili2014binary}. A ideia básica do PSOGSA é combinar a capacidade da busca local da solução presente no PSO com a capacidade de melhor exploração do espaço de busca do GSA. O algoritmo PSOGSA foi modelado matematicamente da seguinte forma: similar ao PSO e o GSA, cada agente de busca tem um vetor de posição refletindo a posição atual nos espaços de busca da seguinte maneira:
\begin{equation}
    \vec{X_i} = (x_i^1, \hdots, x_i^d), \ \ i = 1, 2, \hdots N,
\end{equation}
\noindent em que $N$ é o número de agentes de busca, $d$ é a dimensão do problema e $x_i^1$ é a posição do $i$-ésimo agente na $d$-ésima dimensão.

O processo de otimização se inicia com o preenchimento da matriz de posição com valores aleatórios. Durante o procedimento, a força gravitacional do agente $j$ exercida no agente $i$ em um tempo específico $t$ é definida por:
\begin{equation}
    F_{ij}^{d}(t) = G(t)\frac{M_{pi}(t)\times M_{aj}(t)}{R_{ij}(t)+\epsilon}(x_{j}^{d}(t) - x_{i}^{d}(t)), 
\end{equation}
\noindent em que $M_{aj}$ é a massa gravitacional relacionada ao agente $j$, $M_{pi}$ é a massa gravitacional do agente $i$, $G(t)$ é a constante gravitacional no tempo $t$, $\epsilon$ é uma constante e $R_{ij}(t)$  é a distância Euclidiana entre os agentes $i$ e $j$. As equações para o cálculo de $G(t)$ e $R_{ij}(t)$ são apresentadas em \citep{mirjalili2014binary}. 

A aceleração de cada agente pode ser calculada como \citep{mirjalili2014binary}: 
\begin{equation}
    ac_{i}^{d}(t) = \frac{F_{i}^{d}}{M_{i}(t)'},
\end{equation}
\noindent em que $t$ é um tempo específico e $M_i$ é a massa do agente $i$.

A velocidade e a posição dos agentes são atualizadas respectivamente por:
\begin{align}
    V_{i}(t+1) = w \times v_i(t) + c'_{1} \times rand \times ac_i(t) + \nonumber \\
    c'_2 \times rand \times (gbest -X_i(t)),
\end{align}

\begin{equation}
\label{position}
       X_i(t+1) = X_i(t)+ V_i(t+1).
\end{equation}

\noindent em que $V_i(t)$ é a velocidade do agente $i$ na iteração $t$, $c'_j$ é um coeficiente de aceleração, $w$ é um fator de inércia para controle da velocidade, $rand$ é um número aleatório entre $0$ e $1$, $ac_i(t)$ é a aceleração do agente $i$ na iteração $t$ e $gbest$ é a melhor solução encontrada. Para mapear a probabilidade de mudança de velocidade  e posição em um espaço de busca discreto, utiliza-se uma função hiperbólica definida como:
\begin{equation}
    S(V_{ik}) = \abs{\frac{2}{\pi}\arctan(\frac{\pi}{2}V_{ik})}.
    \label{eq5:tf_v}
\end{equation}

\section{Metodologia}
\subsection{Meta-heurística aplicada à seleção de estruturas NARX polinomiais}
Nesta seção o uso do BPSOGSA para a seleção de estruturas de modelos NARX é proposto. Inicialmente define-se o a dimensão da função de teste ($noV$) e o número de agentes de busca da meta-heurística ($N$), sendo possível gerar uma matriz $noV \times N$ (chamada de $X$) com as respectivas posições dos agentes. O $noV$ tem a dimensão de todos os possíveis regressores a serem incluídos no espaço de busca e cada coluna de $X$ representa uma possível solução. Nessa representação, o vetor binário inicial é definido aleatoriamente e o número $1$ indica que a respectiva coluna é incluída na matriz de regressores reduzida, enquanto $0$ indica que a respectiva coluna é ignorada na solução. Após gerar o modelo candidato, estima-se os parâmetros utilizando o método dos Mínimos Quadrados.

\begin{exemplo}
Considere um caso em que todos os possíveis regressores são definidos baseados em $\ell ~=~1$ e $n_y ~=~ n_u ~=~ 2$. Assim, a matriz de regressores é 
\begin{eqnarray}
[constante \ y(k-1) \ y(k-2) \ u(k-1) \ u(k-2)]
\end{eqnarray}

Nesse caso, tem-se $5$ possíveis regressores, logo $noV = 5$. Se $N ~=~ 5$, a matriz $X$ pode ser representada, por exemplo, como
\begin{equation}
    X = 
    \begin{bmatrix}
0   &  1  &   0  &   0  &   0   \\
1   &  1  &   1  &   0  &   1   \\ 
0   &  0  &   1  &   1  &   0    \\
0   &  1  &   0  &   0  &   1    \\
1   &  0  &   1  &   1  &   0    
\end{bmatrix}
.
\end{equation}

A primeira coluna de $X$ é transposta e usada como uma possível solução. Nesse exemplo, o primeiro modelo testado é $\alpha y_{k-1} + \beta u_{k-2}$, em que $\alpha$ e $\beta$ são parâmetros estimados via MQ. 
\begin{align}
 [constante \ \ \ \ y_{k-1} \ \ \ \  y_{k-2} \ \ \ \ u_{k-1} \ \ \ \ u_{k-2}] \nonumber \\
 \hspace{0.9cm}   0 \hspace{1.5cm}  1 \hspace{1cm} 0\hspace{1.3cm} 0 \hspace{1.3cm} 1 \nonumber
\end{align}

Posteriormente, a segunda coluna de $X$ é testada e assim por diante. Uma vez que todas as colunas de $X$ são testadas, as posições de cada agente de busca em $X$ são atualizadas baseadas na Equação~\eqref{position}. 
\label{exm}
\end{exemplo}

\subsection{Variância dos parâmetros e relevância dos regressores}
A variância dos parâmetros estimados em um modelo pode ser definida como:
\begin{equation}
\label{variancia}
    (\hat{\sigma}_j)^2 = (\hat{\sigma}_e)^2V_{jj},
\end{equation}
\noindent em que $(\hat{\sigma}_e)^2$ é a variância do resíduo estimada, obtida por meio do escalonamento da média quadrática dos resíduos por um fator $N/(N-m)$, e $V_{jj}$ é o $j-$ésimo elemento da diagonal de $(\sum_{k=1}^{N}\psi(k)\psi'(k))^{-1}$. 

A partir da variância dos parâmetros obtida via Equação~\eqref{variancia}, $(\hat{\sigma}_j)^2$, é feito um teste baseado no teste estatístico T de Student para determinar a relevância estatística de cada regressor. Sendo assim, seja $t_{\alpha,N-m}$ a $100(1-\alpha)$ porcentagem da distribuição de penalidade com $N-m$ graus de liberdade, o $100(1-\alpha)\%$ intervalo de confiança de cada $\theta_j$ é dado por \cite{falsone2015randomized}:
\begin{equation}
\label{intervalo}
    [\hat{\theta}_j - \hat{\sigma}_jt_{\alpha,N-m};\hat{\theta}_j + \hat{\sigma}_jt_{\alpha,N-m}].
\end{equation}

Se o intervalo de confiança é definido pela expressão \eqref{intervalo} não contém zero, $\theta_j$ não é zero com um $100(1-\alpha)\%$ de confiança. Caso contrário, $\theta_j$ não é significantemente diferente de zero a hipótese nula $\theta_j = 0$ não pode ser rejeitada. Nesse caso, o regressor correspondente $\Sigma$ é considerado estatisticamente irrelevante para o modelo \citep{van2008estimation}.

\subsection{Função de Penalidade}

Esse trabalho apresenta uma nova abordagem para penalizar o modelo em função da complexidade do mesmo e da contribuição individual de cada regressor ao explicar a resposta do sistema. Para isso, utiliza-se a função Sigmoide e o erro padrão definidos respectivamente como
\begin{equation}
\label{sigmoide}
S(x) = \frac{1}{1+e^{-a(x-c)}},
\end{equation}

\begin{equation}
\label{erro-padrao}
ep(\hat{\Theta}_j) = \sqrt{\sigma^2 C_{jj}},
\end{equation}

\noindent de forma que $x$ tenha a dimensão de $noV$; $c = noV/2$; e $a$ é definido pelo número de regressores que compõem o modelo atual dividido pelo valor $c$. Dependendo do sinal do parâmetro $a$, a função de associação sigmoidal é inerentemente aberta à direita ou à esquerda. Os $(\hat{\Theta}_j)$ são os parâmetros do modelo, $\sigma^2$ é a variância dos resíduos e $C_{jj}$ é o elemento da matriz $\Psi^T\Psi$ que corresponde a $\hat{\Theta}_j$.

\begin{figure}[!ht]
\centering 
 	\includegraphics[width=0.49\textwidth]{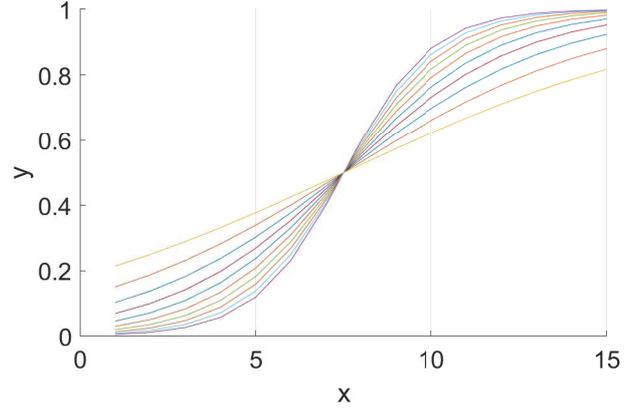}
 	\caption{Diferentes valores da função sigmoide dado o número de termos de diferentes modelos.}
 	\label{sigex}
\end{figure}

Essa abordagem resulta em diferentes curvas geradas pela função Sigmoide baseadas em cada modelo. Uma vez que número de regressores aumenta, a "inclinação"~da curva sigmoide se torna mais íngreme (Figura \ref{sigex}). 

Sabe-se, no entanto, que dois modelos como o mesmo número de regressores podem apresentar resultados muito diferentes. Isso pode ser explicado baseado na importância de cada regressor ao compor o modelo. Nesse sentido, além do número de regressores, é verificado a quantidade de regressores estatisticamente relevantes para o modelo. Para isso, utiliza-se a Equação~\eqref{variancia} para verificar a quantidade de regressores não relevantes para o modelo, $aux$. O valor de $aux$ é somado ao número de regressores do modelo, de forma a penalizar mais aqueles modelos que contém regressores irrelevantes. Dessa forma, mesmo se o os modelos testados possuírem a mesma quantidade de termos, aquele que possui regressores espúrios são mais penalizados no cálculo da função sigmoide.

Sendo assim, o valor de penalidade,~$\rho$, corresponde ao valor em $y$ da curva sigmoide correspondente dado o número de regressores somado ao número de regressores espúrios em $x$.

Portanto, a função custo do problema é definida como 
\begin{equation}
\label{custo}
    f_{custo} = \frac{1}{N}\sum_{k=1}^{N}(y(k)-\hat{y}(k))^2 \times \rho,
\end{equation}
\noindent em que a primeira parcela da equação calcula o erro médio quadrático cometido pelo modelo em simulação livre e $\xi$ é o valor da penalidade a ser aplicada.

\subsection{Estudos de caso}
Para avaliar os resultados obtidos são realizados dois estudos de caso. O primeiro diz respeito à modelagem de um sistema eletromecânico composto por um gerador CC, anteriormente identificado em \citep{junioridentificacao}. O sinal de entrada utilizado nesse sistema foi um sinal binário pseudo-aleatório (PRBS) de tensão e a saída corresponde à velocidade do gerador. Esse caso permite comparar o desempenho do método proposto com a Taxa de Redução de Erro. 
O segundo caso trata da identificação de um aquecedor elétrico utilizando os mesmo dados utilizados por \citep{severino2017meta}. Os dados utilizados foram originalmente coletados de um sistema termo-elétrico, no qual a entrada é um sinal pseudo-aleatório (PRS) de tensão e a saída corresponde à tensão medida nos terminais de um termopar. Isso permite uma comparação direta entre os dois métodos que utilizam meta-heurísticas para seleção de estruturas. Em ambos os casos os dados foram normalizados e divididos em dados identificação e de validação.

A definição dos regressores candidatos é realizada seguindo os mesmos critérios estabelecidos nos artigos utilizados para comparação. Nesse caso, $\ell = n_y = n_u = 2$ para o gerador CC e $\ell =3 $,  $n_y = n_u = 2$ para o aquecedor elétrico. Os demais valores necessário para a execução da meta-heurística são definidos de acordo com a Tabela~\ref{tab:para}.

\begin{table}[ht]
\centering
\caption{Parâmetros utilizados para execução do BPSOGSA}
\label{tab:para}
\resizebox{0.48\textwidth}{!}{%
\begin{tabular}{cccccc}
\hline
Parâmetros & p-value & max\_iter & n\_agentes & $\alpha$ & $G_0$  \\ \hline
Valores    & $0.05$  & $30$      & $10$       & $23$    & $100$ \\ \hline
\end{tabular}%
}
\end{table}

Por fim, é importante mencionar que todas as rotinas são executadas pelo software Matlab\textregistered, em um notebook Dell Inspiron $5448$ Core i$5-5200$U CPU $2.20$GHz com $12$GB de RAM.

\section{Resultados}

\subsection{Gerador CC}
Seguindo a metodologia proposta, o espaço de busca gerado resultou em um total de $2^{15}$ possibilidades de modelos. A Tabela \ref{tab:motor} compara os resultados obtidos utilizando o método proposto com aqueles obtidos via ERR.

\begin{table}[!ht]
\centering
\caption{Comparação dos resultados obtidos para o sistema eletromecânico.}
\label{tab:motor}
\resizebox{0.48\textwidth}{!}{%
\begin{tabular}{cccccc}
\hline
Proposto &  &  & ERR &  &  \\ \hline
Regressor & $\Theta_i$ & MSE & Regressor & $\Theta_i$ & MSE \\
$y_{k-1}$ & 1,8143 &  & $y_{k-1}$ & 1,7823 & \multirow{9}{*}{$1,92\times10^{-4}$} \\
$y_{k-2}$ & -0,8270 & \multirow{6}{*}{$1,37\times10^{-4}$} & $y_{k-2}$ & 0,7969 &  \\
$u_{k-1}$ & 0,0337 &  & $u_{k-1}$ & 0,0337 &  \\
$y_{k-1}^2$ & -1,0292 &  & $y_{k-2}^2$ & 0,0082 &  \\
\multicolumn{1}{l}{$y_{k-1}y_{k-2}$} & \multicolumn{1}{l}{1,9867} &  & \multicolumn{1}{l}{$u_{k-1}y_{k-2}$} & \multicolumn{1}{l}{0,1302} &  \\
\multicolumn{1}{l}{$u_{k-1}y_{k-1}$} & \multicolumn{1}{l}{-0,1917} &  & \multicolumn{1}{l}{$u_{k-1}y_{k-1}$} & \multicolumn{1}{l}{-0,1600} &  \\
\multicolumn{1}{l}{$u_{k-2}y_{k-1}$} & \multicolumn{1}{l}{-0,0292} &  & \multicolumn{1}{l}{$u_{k-2}y_{k-1}$} & \multicolumn{1}{l}{-0,1389} &  \\
\multicolumn{1}{l}{--} & \multicolumn{1}{l}{--} &  & \multicolumn{1}{l}{$u_{k-2}$} & \multicolumn{1}{l}{0,0340} &  \\
\multicolumn{1}{l}{--} & \multicolumn{1}{l}{--} &  & \multicolumn{1}{l}{$u_{k-2}y_{k-2}$} & \multicolumn{1}{l}{0,1077} &  \\ \hline
\end{tabular}%
}
\end{table}

\begin{figure}[!ht]
\centering
        \includegraphics[width=0.48\textwidth]{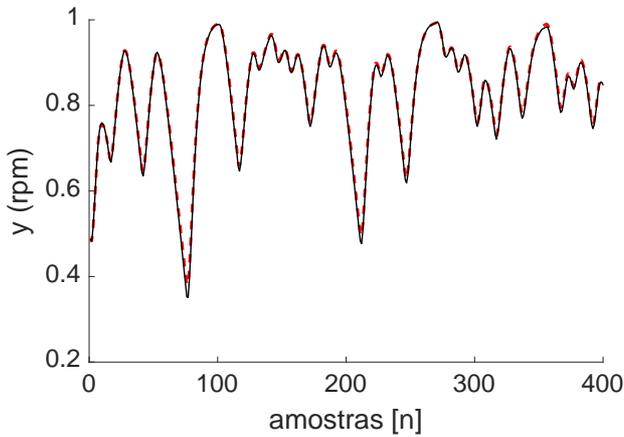}
\caption{Simulação livre do modelo obtido por meio do método proposto. ($-$) são os dados do sistema e ($--$) são dados do modelo.}
\label{fig:motor}
\end{figure}

\begin{figure}[!ht]
\centering
        \includegraphics[width=0.48\textwidth]{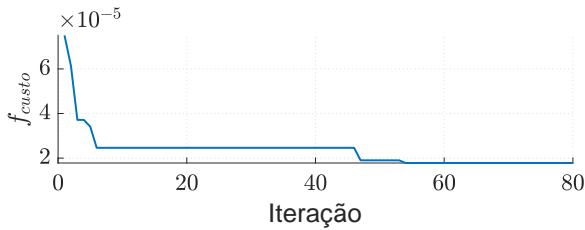}
\caption{Curva de convergência do algoritmo proposto.}
\label{cm}
\end{figure}

Como pode ser observado na Tabela \ref{tab:motor}, o método proposto foi capaz de selecionar um modelo cuja estrutura é mais simples do que a obtida via ERR. Os regressores $u_{k-2}$ e $u_{k-2}y_{k-2}$ não são incluídos na estrutura selecionada pelo método proposto.  Além disso, o MSE alcançado é menor do que o calculado para o modelo selecionado via ERR, o que mostra que a ERR havia determinado uma estrutura mais complexa do que seria necessário para o modelo do sistema. A Figura \ref{fig:motor} mostra o desempenho do modelo ao representar o sistema proposto e a figura \ref{cm} apresenta a curva de convergência do algoritmo.

É importante mencionar que o método proposto é computacionalmente eficiente, gerando o modelo em um tempo $< 1$s, semelhante à técnica ERR. Uma vez que o esforço computacional da ERR é uma das características mais celebradas do algoritmo, o método proposto se mostra uma ferramenta robusta nesse quesito.

\subsection{Aquecedor Elétrico}
o espaço de busca gerado resultou em um total de $2^{35}$ possibilidades de modelos. As Tabelas \ref{tab:aquecedor} e \ref{tab:aquecedorp} comparam os resultados obtidos utilizando o método proposto com aqueles obtidos via meta-heurística em conjunto com AIC. 

\begin{table}[!ht]
\centering
\caption{Regressores selecionados por cada técnica.}
\label{tab:aquecedor}
\resizebox{0.48\textwidth}{!}{%
\begin{tabular}{cccc}
\hline
Proposto & GA    & PSO     & BA                        \\ \hline
$y_{k-1}$  & $y_{k-1}$  & $y_{k-1}$  & $y_{k-1}$ \\
$y_{k-2}$  & $y_{k-2}$  & $y_{k-2}$  & $y_{k-2}$  \\
$u_{k-1}^2$ & $u_{k-2}$ & $u_{k-2}$  & $u_{k-1}$ \\
& $u_{k-1}^2$ & $y_{k-2}u_{k-1}$ & $u_{k-2}$      \\
& $u_{k-1}^2u_{k-2}$ & $y_{k-2}u_{k-2}$ & $y_{k-1}u_{k-2}$  \\
&  & $u_{k-1}^2$ & $y_{k-2}^2$      \\
&  & $u_{k-2}^2$ & $u_{k-1}^2$  \\ \hline
\end{tabular}%
}
\end{table}


\begin{table}[!ht]
\centering
\caption{Parâmetros selecionados e MSE obtido por cada técnica para o aquecedor elétrico.}
\label{tab:aquecedorp}
\resizebox{0.45\textwidth}{!}{%
\begin{tabular}{ccccc}
\hline
     & Proposto                     & GA                          & PSO                          & BA                        \\
     & \multicolumn{1}{c}{$\Theta$} & \multicolumn{1}{c}{$\Theta$}   & \multicolumn{1}{c}{$\Theta$}    & \multicolumn{1}{c}{$\Theta$} \\ \hline
     & $1,2076$                     & $1,4349$                    & $1,4132$                     & $1,4260$                  \\
     & $-0,3037$                    & $-0,4744$                   & $-0,4530$                    & $-0,4621$                 \\
     & $0,0101$                     & $0,0034$                    & $0,0012$                     & $-0,0039$                 \\
     &                              & $0,0115$                    & $0,0295$                     & $0,0047$                  \\
     &                              & $0,0060$                    & $0,0305$                     & $0,0036$                  \\
     &                              &                             & $0,0104$                     & $-0,0186$                 \\
     &                              &                             & $0,0093$                     & $0,0189$                  \\ \hline
MSE & \multicolumn{1}{l}{$7,5627\times10^{-4}$}     & \multicolumn{1}{l}{$1,1084\times10^{-3}$} & \multicolumn{1}{l}{$1,0979\times10^{-3}$} & \multicolumn{1}{l}{$1,1212\times10^{-3}$} \\ \hline
\end{tabular}%
}
\end{table}

\begin{figure}[!ht]
\centering
        \includegraphics[width=0.5\textwidth]{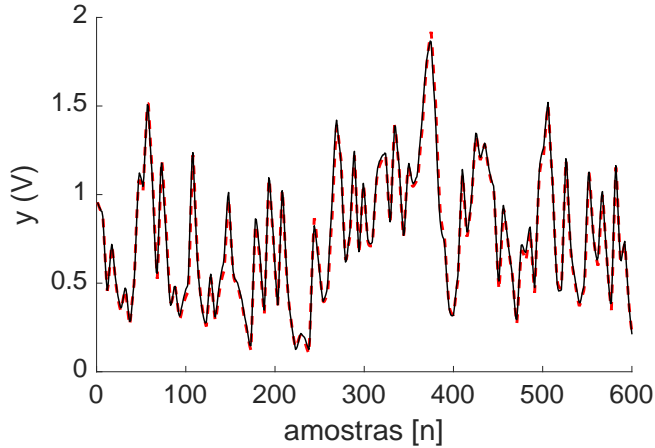}
\caption{Simulação livre do modelo obtido por meio do método proposto. (-) são os dados do sistema e (- \ -) são dados do modelo.}
\label{fig:aquecedor}
\end{figure}

\begin{figure}[!ht]
\centering
        \includegraphics[width=0.48\textwidth]{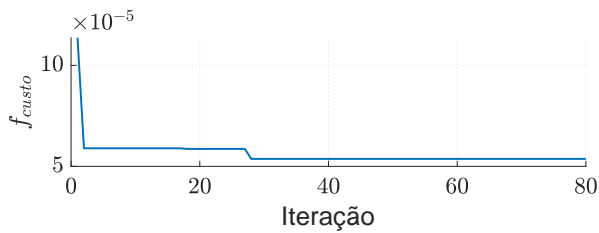}
\caption{Curva de convergência do algoritmo proposto.}
\label{ca}
\end{figure}

Como pode ser observado nas Tabelas \ref{tab:aquecedor} e \ref{tab:aquecedorp}, o método proposto foi capaz de encontrar um modelo com um menor número de regressores e ainda obter um desempenho satisfatório, com um MSE menor do que os obtidos pela técnica utilizada como comparação. É importante ressaltar que o método proposto não impõe nenhuma restrição ao modelo relacionada ao número de termos e valor da função custo, como é feito pela técnica escolhida para comparação. Isso mostra que método proposto se mostra eficaz em selecionar um modelo simples e representativo em uma vasta possibilidade de modelos. A Figura \ref{fig:aquecedor} mostra a predição livre do modelo obtido e a figura \ref{ca} detalha a curva de convergência do algoritmo.

O tempo de execução do BPSOGSA para a obtenção do modelo foi de $1.23$s. Intuitivamente, o aumento no tempo de execução é esperado, uma vez que o espaço de busca aumentou significantemente. No entanto, o esforço computacional continua sendo baixo, principalmente quando comparado à outra técnica baseada em meta-heurísticas: GA $\approx 1188,42 (s)$, PSO $\approx 629,03 (s)$, e BA $\approx 376,02 (s)$. 

\section{Conclusões}
Este artigo apresenta uma nova abordagem para a seleção da estrutura de modelos NARX polinomiais baseada em meta-heurísticas. A esse respeito, uma função de penalidade foi proposta com base na complexidade do modelo (em relação ao número de parâmetros) e na contribuição de cada regressor ao explicar o sistema a ser modelado.

Dois estudos de caso foram realizados para avaliar o desempenho do método. O primeiro caso apresentou dados de um sistema experimental composto por um gerador CC e o segundo dados experimentais de um aquecedor elétrico. Os resultados foram comparados com aqueles obtidos utilizando ERR e meta-heurísticas em conjunto com AIC e restrições. Em relação ao primeiro caso, o método proposto resultou na determinação de um modelo com $2$ regressores a menos e um melhor desempenho.

No segundo caso, o método proposto obteve um modelo com $3$ regressores, enquanto a outra técnica baseada em meta-heurísticas resultou em modelos com $5$ e $7$ regressores. Além disso, o esforço computacional do método proposto é baixo, uma característica relevante em um contexto de identificação de sistemas. 

Por fim, é importante mencionar que a técnica proposta não restringe, como aquela que está sendo comparada, a ordem do modelo ou o valor máximo da função custo a ser minimizada. Isso, em um contexto em que não se tem informação a respeito do modelo, faz da método proposto uma alternativa eficaz em situações práticas.

\section*{Agradecimentos}
Os autores agradecem ao CNPq/INERGE, FAPEMIG e CAPES pelo apoio financeiro.

\bibliography{ifacconf}             
                                                   







\end{document}